\documentclass{article}
\usepackage{spconf,amsmath,graphicx}
\usepackage{xcolor} %
\usepackage{color}
\usepackage{multirow}
\usepackage{booktabs}
\usepackage{tabularx}
\usepackage[symbol]{footmisc}
\usepackage{hyperref}

\title{Futga: Towards Fine-grained Music Understanding through Temporally-enhanced Generative Augmentation}

\name{
\begin{tabular}{@{}c@{}}
Junda Wu$^{1*}$, 
Zachary Novack$^{1*}$, 
Amit Namburi$^1$, 
Jiaheng Dai$^1$ \\
Hao-Wen Dong$^1$, 
Zhouhang Xie$^1$,
Carol Chen$^2$,
Julian McAuley$^1$ \\ 
\end{tabular}
}
\address{$^1$University of California, San Diego, $^2$University of California, Los Angeles}
 
\begin{document}
\maketitle

\begin{abstract}
Existing music captioning methods are limited to generating concise global descriptions of short music clips,
which fail to capture fine-grained musical characteristics and time-aware musical changes.
To address these limitations, we propose FUTGA, 
a model equipped with fined-grained music understanding capabilities through learning from generative augmentation with temporal compositions.
We leverage existing music caption datasets and large language models (LLMs) 
to synthesize fine-grained music captions with structural descriptions and time boundaries for full-length songs.
Augmented by the proposed synthetic dataset, FUTGA is enabled to identify the music's temporal changes at key transition points and their musical functions, 
as well as generate detailed descriptions for each music segment.
We further introduce a full-length music caption dataset generated by FUTGA, as the augmentation of the MusicCaps and the Song Describer datasets.
We evaluate the automatically generated captions on several downstream tasks, including music generation and retrieval.
The experiments demonstrate the quality of the generated captions and the better performance in various downstream tasks achieved by the proposed music captioning approach. 
Our code and datasets can be found in \href{https://huggingface.co/JoshuaW1997/FUTGA}{\textcolor{blue}{https://huggingface.co/JoshuaW1997/FUTGA}}.
\end{abstract}

\section{Introduction}\label{sec:introduction}
\begin{figure}[h!]
    \centering
    \includegraphics[width=1.\linewidth]{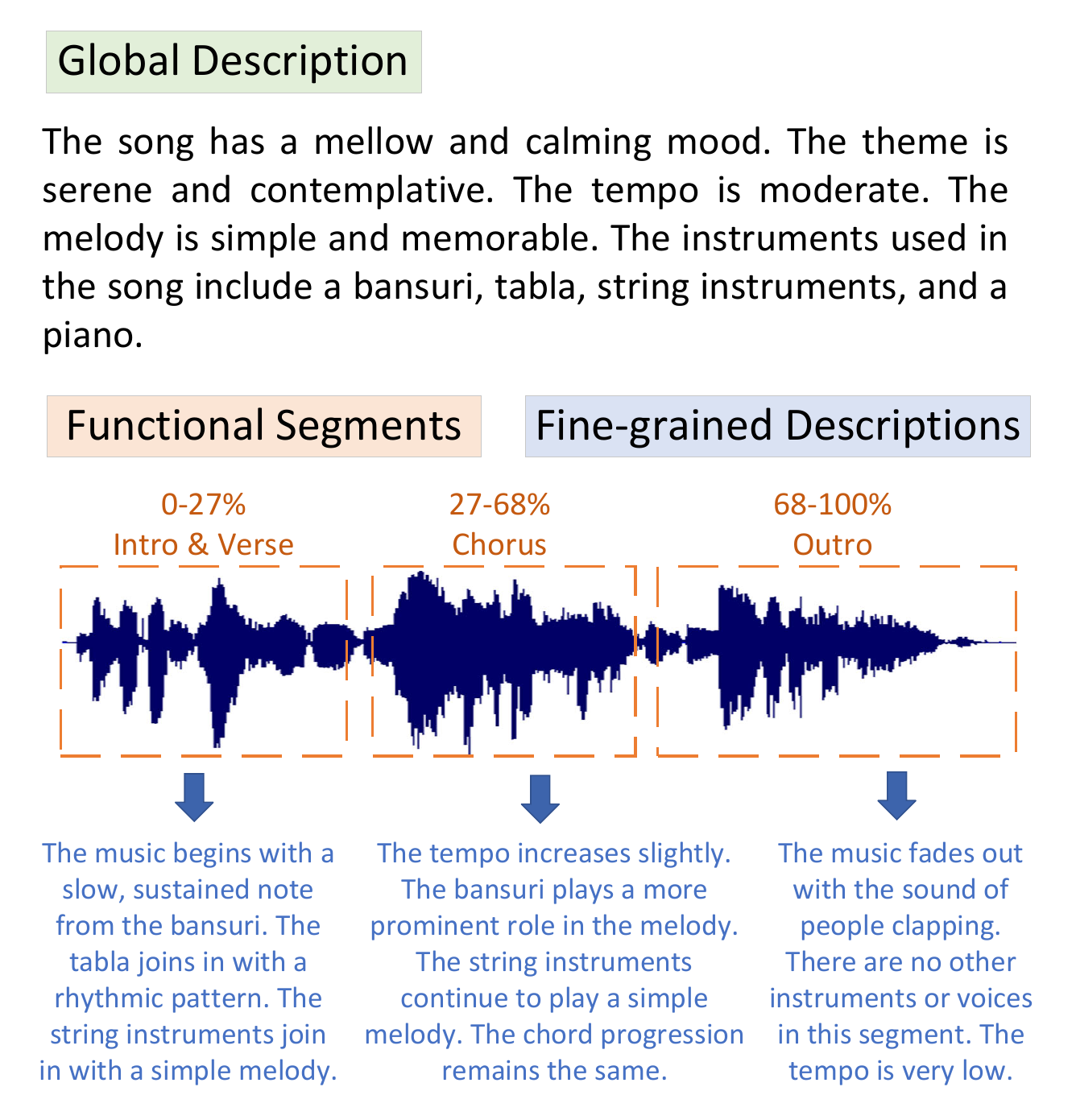}
    \caption{\textbf{Overview of FUTGA's capabilities.} Given a long-form audio example, FUTGA is able to provide time-located captions by automatically detecting functional segment boundaries, as well as global captions.}
    \label{fig:enter-label}
\end{figure}
Natural language music understanding, which extracts music information and generates detailed music captions,
is a fundamental task within the MIR community, beneficial for a series of applications including 
music generation \cite{copet2024simple, chen2024musicldm,novack2024ditto,melechovsky2023mustango}, editing \cite{wang2023audit,zhang2024musicmagus}, 
question-answering \cite{deng2023musilingo,gao2022music}, and retrieval \cite{doh2023toward,wu2023clamp,bhargav2023music}.
Existing music understanding and captioning methods focus mainly on model architecture design and data augmentation.
Traditional music understanding models are developed for specific purposes and with dedicated model design, 
including single-track music captioning \cite{cai2020music,manco2021muscaps,doh2023lp}, 
playlist title generation \cite{doh2021music} and conversational music recommendation \cite{leszczynski2023talk,dong2023musechat}.
Recent developments in music foundation models \cite{gardner2023llark,hussain2023m,tang2023salmonn,liu2024music} enable free-form music prompts and multitasking.
These foundation models are developed based on pre-trained large language models (LLMs) and aligned with the music modality.
Although LLM-powered music understanding models can leverage the abundant pre-trained music knowledge in caption generation,
the success of modality alignment still requires a large amount of high-quality music caption data.

Many previous works have tried to enhance models' music understanding capacity by expanding and improving music annotations on a broader range of music audio files \cite{manco2023song,manco2021muscaps},
or leveraging more powerful large language models to augment the existing music caption datasets with better language quality \cite{doh2023lp,huang2023noise2music} via auto-tagging.
However, restricted by the current music captioning paradigm, available music caption datasets are limited to two major challenges:
(1) Conventional music captions focus only on the global description of a (potentially long) music clip,
which cannot efficiently capture a piece of music's fine-grained characteristics nor differentiate it from other music within-genre songs.
(2) Key structural information, such as time boundaries of functional music segments and time-aware musical changes,
is mostly neglected in traditional music understanding and hard to retrieve due to the limitation in the length of music clips.

To address the limitations, we propose \textbf{FUTGA}, 
a generative music understanding model trained with time-aware music caption data and calibrated with Music Information Retrieval (MIR) features.
We first augment the MusicCaps dataset \cite{agostinelli2023musiclm} by mixing music clips together into synthetic full-length songs.
The corresponding music captions are composed with original short music captions as individual segment descriptions,
which are also tagged with temporal segmentation information.
To enable more realistic full-length music captioning, we further leverage a text-only LLM for the augmentation of 
the global music caption, musical changes between segments (\emph{e.g.}, increase of volume, slowing down the tempo, introducing new instruments, etc.), 
and functional tags of the segments (\emph{e.g.}, intro, verse, chorus, etc.), by paraphrasing and summarizing the template-based captions.

Inspired by existing
Large Audio-Language Models (LALMs), 
we use the open-source SALMONN model \cite{tang2023salmonn} as the backbone
and fine-tune the model with our developed synthetic full-length music caption data.
Using our synthetic data augmentation,
FUTGA is able to identify key transition points in musical changes and segment full-length songs according to their musical functions.
For example, in Figure 1, we illustrate FUTGA's capacities as a novel form of music captioning.
Given a song in full length, FUTGA can generate a global caption that summarizes the whole song's characteristics before identifying the music structure with time segments.
Following the flow of music structures, FUTGA can further describe each music clip and musical changes between consecutive music clips. 
In addition, we also discover that the fine-tuned SALMONN model demonstrates a great instruction-following capacity to generate fine-grained music captions conditioned on given time boundaries and MIR features.
Empowered by such model capacities, the model can further be trained with a small-scale MIR dataset Harmonixset \cite{nieto2019harmonix}, where music structure information and some MIR features are available.
By injecting the ground-truth information into the instruction prompt, we can accurately guide the model to generate fine-grained music captions corresponding to the time segments.
Finally, we ask human annotators to revise the generated captions and fine-tune the model again, which can additionally reduce the sim-to-real gap between synthetic and realistic data distributions.

With the final version of FUTGA, we propose automatically annotating the full-length songs in two existing datasets MusicCaps \cite{agostinelli2023musiclm} and Song Describer \cite{manco2023song}.
We evaluate the effectiveness of our fine-grained and temporal-enhanced music caption in several downstream tasks including music captioning, music retrieval, and music generation.
We summarize our main contributions as follows:
\begin{itemize}
    \item We propose a synthetic data augmentation method to construct fine-grained and temporally-structured music captions for full-length songs.
    \item We fine-tune the existing large audio-language model with the synthetic dataset, 
    and demonstrate the model's emerging ability in music segmentation and fine-grained music understanding.
    \item Further aligned with human annotations and MIR features, 
    the music understanding model FUTGA is used to automatically augment the MusicCaps and the Song Describer datasets for full-length music captions.
    \item Through evaluation experiments, we demonstrate the proposed music captioning paradigm can benefit several downstream music understanding tasks.
\end{itemize}

\section{Temporally-enhanced Generative Augmentation}
In this section, we introduce our proposed temporally-enhanced generative augmentation.
Due to the limitation of existing music caption datasets, music captioning and understanding models can only generate global music descriptions for short music clips \cite{manco2023song,agostinelli2023musiclm,doh2023lp}.
To address this limitation, we propose the augmentation of synthetic music and caption composition, 
which empowers music understanding models with capacities of time-aware music segmentation and fine-grained music description generation. 

\subsection{Synthetic Music Caption Augmentation}
We construct our synthetic music caption augmentation from the existing MusicCaps dataset \cite{agostinelli2023musiclm}.
Due to the limitation of the length (10 seconds) of the music clips in MusicCaps, the annotated captions only capture the global characteristics of the music,
mostly neglecting fine-grained music information, such as musical structures, temporal musical changes and time-boundaries for specific musical segments.
To augment MusicCaps and enable fine-grained music understanding with temporally-enhanced music information,
we propose a synthetic music composition method that composes multiple music clips into a synthetic piece of music with comparable lengths of complete songs.
Given the existing $N$ music clips $C=\{c_i\}_{i=1}^N$ and their corresponding music captions $T=\{t_i\}_{i=1}^N$,
the synthetic music composition method aims to sample a subset of music clips $C_k \subset C$ and the corresponding subset of music captions $T_k \subset T$.
However, for the music parts in a single song, randomly sampling music clips with heterogeneous musical characteristics may produce incoherent music,
which further causes the out-of-distribution (OOD) problem of the constructed synthetic dataset.

To address such music incoherent problems in synthetic music samples and make the constructed music and song descriptions more realistic,
we propose an important sampling method based on the music characteristic distribution.
We extract the semantic embeddings of the original music captions with a pre-trained sentence embedding model $f$,
\begin{equation*}
    \boldsymbol{z}_i = f(t_i), \quad i=1,2,\dots,N,
\end{equation*}
in which $z_i\in \mathcal{R}^{384}$ and we adopt the MiniLM-L12 model \cite{reimers2019sentence} that allows the maximum of $256$ tokens.
With the extracted sentence embeddings, we can estimate the music description similarity by calculating the pairwise cosine similarity,
\begin{equation*}
    d(\boldsymbol{z_i}, \boldsymbol{z_j}) = \frac{\boldsymbol{z_i}^\top \boldsymbol{z_j}}{\|\boldsymbol{z_i}\|\cdot\|\boldsymbol{z_j}\|},
\end{equation*}
which can be used as the important scores for relevant music clip sampling.
Then, for each music clip $c_k\in C$, we use it as the seed sample to explore $n$ relevant music clips for synthetic composition,
\begin{align*}
    C_k &= \{c_{k,1},c_{k,2},\dots,c_{k,n}\}, \\
    c_{k,j} &\sim P(j)=\frac{\exp d(\boldsymbol{z_k}, \boldsymbol{z_j})}{\sum_i \exp d(\boldsymbol{z_k}, \boldsymbol{z_i})},
\end{align*}
in which the number of music clips is uniformly sampled $n\sim U(3,5)$, which is consistent to most music structures.
By sampling from the neighborhood of each music clip, we obtain a set of synthetic whole-songs $\Tilde{C}=\{C_k\}_{k=1}^N$, 
which can both cover the entire music representation space and approximate the representation distribution of realistic songs. 
\begin{figure}[htbp]
    \centering
    \includegraphics[width=0.4\textwidth]{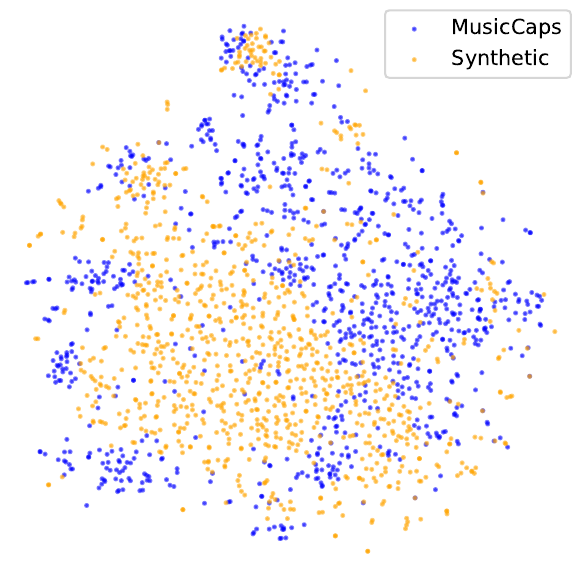}
    \caption{T-SNE representation distribution of the mixed samples of original MusicCaps songs and synthetic songs.}
    \label{fig:tsne}
\end{figure}
We visualize the joint distribution of the original MusicCaps and the synthetic music representations in Figure \ref{fig:tsne}.
We observe that the distribution of synthetic samples is perfectly aligned with that of the original MusicCaps.
In addition, by augmenting the original dataset, we can bootstrap synthetic samples that are more equally distributed in the feature space, which helps to smooth the learning process.
To promote the time variance of different music segments in different songs, we propose to further modify the clip lengths $L_k=\{l_j\}_{j=1}^n$ with the uniform distribution $l_j \sim U(6,10)$,
which can efficiently reduce the model's bias in music segmentation and promote the model to discover the accurate transition points between music segments.

\subsection{Temporally-enhanced Music Understanding}
Corresponding to synthetic music augmentation, we also augment the music captions with temporally-enhanced music understanding.
For each sampled set of music clips $C_k$, the corresponding music caption set $T_k$ and the clip length information $L_k$ are interleaved and composed by the template,
\begin{align*}
    \Tilde{X}_k = \left\{ (\frac{l_{k,j-1}}{\sum_i l_{k,i}}, \frac{l_{k,j}}{\sum_i l_{k,i}}, t_{k,j}) \right\}_{j=1}^n,
\end{align*}
in which the  specific original time-boundaries $L_k$ are transformed into relative time-boundaries, which are always between $0$ -- $100\%$ ($l_{k,0}\equiv 0$).
We use a relative time-boundary representation approach to minimize training bias towards specific numbers of music lengths in our model.
In addition, relative time-boundary representation enhances the model's ability to comprehend music of varying lengths, thereby improving the model's generalizability.
By generating these relative time boundaries, our generative music understanding model gains a better awareness of the music's overall progression, which further enhances the model's temporal understanding of music.

We further propose to use a text-only large language model (LLM) to augment the template-based caption $\Tilde{X}_k$ with natural language descriptions,
in which additional information, such as global captions, musical changes, and music structures, can be automatically extracted from the LLM.
Since LLMs are pre-trained with abundant domain knowledge including music analysis \cite{tang2023salmonn,gardner2023llark} and music information retrieval \cite{tang2023salmonn,hussain2023m},
with enough context provided, such LLMs can accurately extract music information via language-based summarization and reasoning \cite{doh2023lp}.
Inspired by LLMs' capacities in language reasoning, we propose to paraphrase and augment additional music information with instructions as follows:

\vspace{.5em}
\noindent\textbf{Context}: 
\textit{Music Analysis} \{$\Tilde{X}_k$\}.
\textit{This is a music analysis of a song. Note that the numbers indicate the time-boundaries of functional segments in this song.}

\vspace{.5em}
\noindent\textbf{Paraphrase}: \textit{Paraphrase the music analysis to make it sound like a coherent song, instead of a remix. Additionally, remove any mention of sound quality.}

\vspace{.5em}
\noindent\textbf{Global Caption}: \textit{Start with a general description of the song focusing on subjectivity.}

\vspace{.5em}
\noindent\textbf{Musical Change}: \textit{Describe the song in detail and explain transitions between parts of the song.}

\vspace{.5em}
\noindent\textbf{Music Structure}: \textit{Remember to indicate the temporal annotations and music structures when talking about a specific part of the song.}
\vspace{.5em}

We obtain $5$K synthetic training samples by prompting the GEMMA-7B model \cite{team2024gemma} with the template-based caption $\Tilde{X}_k $ and the designed instructions.
Then we adopt LoRA \cite{hu2021lora} instruction finetuing of the SALMONN-7B \cite{tang2023salmonn} backbone model for 100 epochs and the learning rate of 1e-5, with 2 NVIDIA RTX A6000 GPUs with 48GB each.
We use the bfloat16 type for training with the batch size set to 4 and gradient accumulation steps to 8.

\section{Alignment from MIR features and human feedback}
\begin{table*}[htbp]
\centering
\small
\begin{tabularx}{\textwidth}{X}
\toprule

\textbf{MusicCaps}: 
The low quality recording features footsteps, birds chirping, distant traffic and wind sound effects. At the very end, there is a suspenseful, low, widely spread pad chord. It sounds like it is a sound for some kind of video. \\

\midrule

\textbf{Song Describer}: 
A rock song with a French speaking male voice. It has the typical feeling of political/social rock song with a really simple guitar, bass and drum skeleton. \\

\midrule

\textbf{LP-MusicCaps}: 
The low quality recording features a drum solo that consists of a double pedal kick, punchy snare, shimmering cymbals and boomy toms. It sounds energetic and manic, thanks to that kick pattern. The recording is mono and noisy. \\

\midrule

\textbf{FUTGA}: 
The song exhibits a diverse range of musical styles and instrumentation. Overall, the mood is upbeat and energetic, with a theme of youthfulness and optimism. The tempo is moderate to fast, with a catchy melody and a simple chord progression.\\
\textbf{0-27\%: Intro and Verse} - The song begins with a haunting intro featuring syncopated rhythms and atmospheric chords. The verse introduces the main melody, which is characterized by a catchy hook and a simple chord progression.\\
\textbf{27-58\%: Chorus and Bridge} - The chorus features a soaring melody and energetic vocals. The bridge introduces a new melody and chord progression, which contrasts with the verse and chorus.\\
\textbf{58-79\%: Rock Section} - The song transitions into a rock section with a distorted electric guitar and a heavy drum beat. The section has a groovy feel and could be used in the soundtrack of a high school drama TV series.\\
\textbf{79-100\%: Acoustic Ballad} - The song concludes with an acoustic ballad, sung in a melancholic manner. The melody is simple and the chord progression is repetitive. \\

\bottomrule

\caption{A comparison example of captions generated or annotated by MusicCaps \cite{agostinelli2023musiclm}, Song Describer \cite{manco2023song}, LP-MusicCaps \cite{doh2023lp} and FUTGA.}
\label{tab:caption_comparison}

\end{tabularx}
\end{table*}

\begin{table*}[htb]
\centering
\small

\begin{tabular}{lcccccccc}
\toprule
Dataset 
& \multicolumn{1}{c}{\# Caption} 
& \multicolumn{1}{c}{\# Segment} 
& \multicolumn{1}{c}{Tokens} 
& \multicolumn{1}{c}{Vocab.} 
& \multicolumn{1}{c}{\# Inst.} 
& \multicolumn{1}{c}{\# Genre} 
& \multicolumn{1}{c}{\# Mood} \\ 

\midrule
MusicCaps       & 6k   &  -- & $48.9\pm17.3$ & 6,144 &  75&  267&  146 \\ %
Song Describer  & 1k   &  -- & 21.7$\pm$12.4 & 2,859 &  39&  152&  122 \\ %
LP-MusicCaps    & 542k & --  & 45.3$\pm$28.0 & 1,686 &  65&  239&  151 \\ %
FUTGA           & 7k   & 4.32  &  472.419$\pm$88.5   & 3,537  &  64&  187&  128 \\ %

\bottomrule
\end{tabular}

\caption{Statistics summarization of generated or annotated music captions of baselines and FUTGA.}
\label{tab:stat}
\end{table*}

In this section, we design quality control methods for the generative music understanding model. 
Although the model trained with the synthetic music and captions is capable of generating fine-grained music captions with time-boundary information,
the generation quality on realistic whole-songs remains to be evaluated, especially when a sim-to-real domain gap exists.
To further validate the generation quality of our proposed model and align the model generation distribution with realistic music samples,
we propose to collect a small portion of human annotated ground-truth music captions based on the Harmonixset dataset \cite{nieto2019harmonix}.

In the Harmonixset dataset, there are 912 songs annotated with music structures and MIR features (\emph{e.g.}, genre and BPM).
We obtain the pseudo-labeled music captions by incorporating human annotations into the prompts:

\vspace{.5em}
\noindent\textbf{Pseudo-label}: \textit{This is a} \{\textbf{genre}\} \textit{music of} \{\textbf{bpm}\} \textit{beat-per-minute (BPM).} 
\textit{Describe the music in general, in terms of mood, theme, tempo, melody, instruments, and chord progression. }
\textit{Then provide a detailed music analysis by describing each functional segment and its time boundary. Please note that the music boundaries are } \{\textbf{segments}\}.
\vspace{.5em}

\noindent We observe that with the time segments provided in the prompts, the model is capable of accurately following such requirements.
In addition, we ask four human annotators to inspect in detail the generated music descriptions and correct description errors based on their listening to the specific music segments.
With the ground-truth labels, we fine-tune the previously trained FUTGA model again with a relatively smaller learning rate of 1e-6 and the training epochs of 20, to prevent the model from overfitting.

\section{Dataset Creation: FUTGA}
\begin{table*}[htp]
\centering
\small

\begin{tabular}{lcccccccccccc}
\toprule

 & \multicolumn{6}{c}{MusicCaps} & \multicolumn{6}{c}{Song Describer} \\
\cmidrule(lr){2-7}                      \cmidrule(lr){8-13}                  
Model & B1 & B2 & B3 & M & R & B-S & B1 & B2 & B3 & M & R & B-S \\ 
 
\midrule
LP-MusicCaps & 19.77 & 6.70 & 2.17 & 12.88 & 13.03 & 84.51 & 1.68 & 0.71 & 0.27 & 7.68 & 2.76 & 79.62\\
FUTGA (complete) & 9.21 & 4.18 & 1.97 & \textbf{20.85 }& 11.96 & 82.62 & 4.58 & 1.72 & 0.61 & 12.82 & 6.90 & 81.21 \\
FUTGA (global) & \textbf{26.46} & \textbf{10.93} & \textbf{4.66} & 18.60 & \textbf{17.40} & \textbf{86.48} & \textbf{14.23} & \textbf{5.04} & \textbf{1.75} & \textbf{15.04} & \textbf{11.67} & \textbf{85.42} \\

\bottomrule

\end{tabular}

\caption{Comparison results of caption generation for LP-MusicCaps and FUTGA.}
\label{tab:cap}

\end{table*}

\begin{table*}[htp]
\centering
\small
\begin{tabular}{lccccccccc}
\toprule
 &  \multicolumn{4}{c}{MusicCaps} & \multicolumn{4}{c}{Song Describer} \\
\cmidrule(lr){2-5}                      \cmidrule(lr){6-9}                  
\multirow{-1}{*}{Caption Model}  & R@1 & R@5 & R@10 & MedR & R@1 & R@5 & R@10 & MedR \\ 

\midrule
\multirow{-1}{*}{Human Annotation}  & \textbf{4.83} & \textbf{14.05} & \textbf{20.83} & \textbf{59} & 4.42 & 17.02 & 26.01 & 36 \\

\multirow{-1}{*}{LP-MusicCaps}  & 3.13 & 11.36 & 16.81 & 79 & 0.18 & 0.88 & 1.77 & 283 \\

\multirow{-1}{*}{FUTGA}         & 3.30 & 8.49 & 13.21 & 160 & \textbf{6.15} & \textbf{18.17} & \textbf{27.61} & \textbf{33}  \\ 

\bottomrule
\end{tabular}

\caption{Comparison results for music retrieval with human annotated captions, LP-MusicCaps and FUTGA.}
\label{tab:retrieval}
\end{table*}

Based on the final version of our proposed music captioning model, 
we automatically generate music captions for whole songs between 2 minutes and 5 minutes in MusicCaps \cite{agostinelli2023musiclm} and Song Describer \cite{manco2023song}.
During inference time, we set the repetition penalty as 1.5 to prevent repetitive descriptions of the same music segments.
In addition, we also set the beam search number to 10 to find the statistically best captions.
We allow a maximum of 2048 tokens to be generated from FUTGA.

As demonstrated in the comparison example in Table \ref{tab:caption_comparison}, 
FUTGA provides more fine-grained music understanding descriptions with time boundaries indicating music segments,
for which the average segment number and the number of musical changes are reported in Table \ref{tab:stat}.
In addition, we can observe relatively longer global captions with more details, which is also verified by the data statistics in Table \ref{tab:stat}. 

In terms of music caption diversity, we first show that our captions have significantly larger numbers of tokens and vocabulary size, compared to existing music caption datasets.
Second, our dataset still maintains good diversity in terms of unique genre, instrument, and music mood vocabularies, which are comparable to human or GPT-3.5 annotations. 
Thus, the FUTGA dataset can serve to augment existing music captioning models with strong temporal reasoning abilities without harming the model's generalizability,
which will be further evaluated in our evaluation section. 

\section{Automatic Music Evaluation}

\subsection{Caption Generation}
We first evaluate the generated data samples' quality by comparing them to existing human annotation datasets, MusicCaps \cite{agostinelli2023musiclm} and Song Describer \cite{manco2023song}.
We follow the previous works \cite{doh2023lp, manco2023song} and report the metrics, BLEU (B), METEOR (M), ROUGE (R), and BERT-score (B-S), in Table \ref{tab:cap}. 
Since our captions are formally different from original music captions, we report the evaluation metrics for the global and the complete captions in our dataset separately.
For a fair comparison, we adopt the zero-shot performance of LP-MusicCaps in \cite{doh2023lp}, since our model is only trained on the synthetic dataset and Harmonixset.

Based on the results in Table \ref{tab:cap} on MusicCaps, 
we observe that the global captions generated from our model consistently show higher quality than the zero-shot results of LP-MusicCaps,
which demonstrates that by capturing more details from longer songs, we can obtain more accurate descriptions of the music.
In addition, comparing FUTGA and LP-MusicCaps on Song Describer, which is the out-of-domain dataset for both methods, 
FUTGA shows a significantly larger improvement in the generation results, which demonstrates the model's better capacities in generalizability. 

However, the complete music captions generated from FUTGA show relatively inferior performance on MusicCaps,
which is mainly due to the different forms of music captions. 
Since FUTGA focuses on the temporal reasoning of a whole song, the time segment information and musical changes are completely new to both the original MusicCaps captions.
Whereas, LP-MusicCaps is directly augmented from MusicCaps, which makes their captions formally more similar.
Such observations can motivate future works to explore more fine-grained and complex music caption forms in terms of evaluating the model's generation capacities.

\subsection{Music Retrieval}
With segmented music descriptions enabled by FUTGA, we propose a many-to-many music retrieval method by matching multiple music segment descriptions with multiple audio clips.
Specifically, we first calculate the intersection-over-union (IoU) score of each music caption's time segment and each audio clip's time segment, 
which is regarded as the pairwise temporal correlation weight between $0$ and $1$.
Then, we adopt the contrastive multimodal representation model CLAP \cite{wu2023large} to extract the text and audio features for each caption and audio segment.
Finally, we calculate the IoU weighted average cosine similarity between all the music captions and audio clips.
Intuitively, the more temporally aligned caption-audio pairs are gaining more importance scores, whereas non-overlapped pairs will be regarded as irrelevant.

In Table \ref{tab:retrieval}, we follow the previous works \cite{doh2023lp,manco2023song} and report the Recall@K scores, where $K=1,5,10$, as well as median rank (MedR).
On Song Describer, we observe a similar generalization problem across the domain of the LP-MusicCpas baseline,
which shows significantly lower performance compared with the original human annotations.
Whereas, our method achieves better retrieval results than the human annotation baseline, 
which demonstrates the effectiveness of such a many-to-many retrieval method enabled by our fine-grained music descriptions.

On MusicCaps, we observe a consistent performance drop for both LP-MusicCaps and FUTGA, compared with human-annotated captions. 
Since LP-MusicCaps and our methods are augmented by LLMs that have a relatively much more complex language variance compared to CLAP text encoders,
we argue that the complexity of language introduced by LLMs may be out-of-distribution of CLAP, which further deteriorates the retrieval performance.

\subsection{Music Generation}
\begin{table}
    \centering
    \small
    \begin{tabular}{lc|c}
    \toprule
         Model&  Caption& CLAP\\
    \midrule
         MusicLDM &  Ground-Truth& \textbf{0.3825}\\
         MusicLDM-F&  Ground-Truth& 0.3820\\
         \midrule
         MusicLDM&  FUTGA&  0.3225\\
         MusicLDM-F&  FUTGA& \textbf{0.3559}\\
    \bottomrule
    \end{tabular}
    \caption{Music Generation results on ground-truth and FUTGA-generated captions from SongDescriber, where MusicLDM-F is finetuned on MusicCaps with FUTGA captions.}
    \label{tab:gen}
\end{table}

 As most modern text-music datasets include global captions for long audio segments, text-to-music (TTM) generation systems face a ``many-to-one'' problem \cite{chen2024musicldm, zhuo2023video, melechovsky2023mustango}, where overall text relevance may suffer due to learning correlations between global text prompts and weakly related audio segments. Thus, In order to validate FUTGA's ability to provide high-quality time-located audio captions, we propose a music generation  task using the open source music generation model MusicLDM \cite{chen2024musicldm}. Specifically, we measure the CLAP score \cite{wu2023large}, which measures audio-text similarity, between MusicLDM generations using both the original and FUTGA-generated captions from SongDescriber \cite{manco2023song}. Additionally, to assess whether FUTGA can be used as a method to improve TTM training through improving caption data, we finetune MusicLDM on the FUTGA-generated captions for MusicCaps (which we denote MusicLDM-F), using the standard $\epsilon$-prediction loss \cite{song2020score} with a learning rate of 1e-6 for 2000 iterations (see \cite{chen2024musicldm} for more details).

 We show results in Tab.~\ref{tab:gen}. We find that without finetuning, MusicLDM's text relevance suffers when provided with more local FUTGA captions compared to global ones. By finetuning on time-aware FUTGA captions however, MusicLDM-F is able to noticeably improve its ability to follow fine-grained captions while sacrificing little ability to handle global controls.

\section{Conclusion}
In this work, we propose a temporally-enhanced music caption augmentation method through generative large language models. 
By bootstrapping existing music captions with time boundary tags, MIR features, and musical changes,
we fine-tune the pre-trained music understanding model SALMONN-7B, where we observe emerging music segmentation capacities and enable instruction prompting to guide the generation with ground-truth time segments.
To further reduce the sim-to-real gap and align the model with realistic music characteristics, 
we collect human annotations and fine-tune the model again.
We use the fine-tuned model to re-annotate the existing MusicCaps and Song Describer datasets with full-length songs.
The generated captions are shown to be more fine-grained and beneficial for various downstream tasks.

For future works, since our model is the first to enable end-to-end full-length song captioning with significantly longer context provided (10 times more than conventional music captions),
we are motivated to further develop a long-context-based CLAP model, which can enable more complex and longer music retrieval tasks.
In addition, with more fine-grained details provided by our captions, we propose to further use such captions for more complex music understanding tasks, including music question-answering and whole-song generation.

\bibliographystyle{IEEEbib}
\bibliography{Template}

\begin{thebibliography}{10}

\bibitem{copet2024simple}
Jade Copet, Felix Kreuk, Itai Gat, Tal Remez, David Kant, Gabriel Synnaeve,
  Yossi Adi, and Alexandre D{\'e}fossez,
\newblock ``Simple and controllable music generation,''
\newblock {\em Advances in Neural Information Processing Systems}, vol. 36,
  2024.

\bibitem{chen2024musicldm}
Ke~Chen, Yusong Wu, Haohe Liu, Marianna Nezhurina, Taylor Berg-Kirkpatrick, and
  Shlomo Dubnov,
\newblock ``Musicldm: Enhancing novelty in text-to-music generation using
  beat-synchronous mixup strategies,''
\newblock in {\em ICASSP 2024-2024 IEEE International Conference on Acoustics,
  Speech and Signal Processing (ICASSP)}. IEEE, 2024, pp. 1206--1210.

\bibitem{novack2024ditto}
Zachary Novack, Julian McAuley, Taylor Berg-Kirkpatrick, and Nicholas~J Bryan,
\newblock ``Ditto: Diffusion inference-time t-optimization for music
  generation,''
\newblock {\em arXiv preprint arXiv:2401.12179}, 2024.

\bibitem{melechovsky2023mustango}
Jan Melechovsky, Zixun Guo, Deepanway Ghosal, Navonil Majumder, Dorien
  Herremans, and Soujanya Poria,
\newblock ``Mustango: Toward controllable text-to-music generation,''
\newblock {\em arXiv preprint arXiv:2311.08355}, 2023.

\bibitem{wang2023audit}
Yuancheng Wang, Zeqian Ju, Xu~Tan, Lei He, Zhizheng Wu, Jiang Bian, et~al.,
\newblock ``Audit: Audio editing by following instructions with latent
  diffusion models,''
\newblock {\em Advances in Neural Information Processing Systems}, vol. 36, pp.
  71340--71357, 2023.

\bibitem{zhang2024musicmagus}
Yixiao Zhang, Yukara Ikemiya, Gus Xia, Naoki Murata, Marco Mart{\'\i}nez,
  Wei-Hsiang Liao, Yuki Mitsufuji, and Simon Dixon,
\newblock ``Musicmagus: Zero-shot text-to-music editing via diffusion models,''
\newblock {\em arXiv preprint arXiv:2402.06178}, 2024.

\bibitem{deng2023musilingo}
Zihao Deng, Yinghao Ma, Yudong Liu, Rongchen Guo, Ge~Zhang, Wenhu Chen, Wenhao
  Huang, and Emmanouil Benetos,
\newblock ``Musilingo: Bridging music and text with pre-trained language models
  for music captioning and query response,''
\newblock {\em arXiv preprint arXiv:2309.08730}, 2023.

\bibitem{gao2022music}
Wenhao Gao, Xiaobing Li, Cong Jin, and Yun Tie,
\newblock ``Music question answering: Cognize and perceive music,''
\newblock in {\em 2022 IEEE International Conference on Multimedia and Expo
  Workshops (ICMEW)}. IEEE, 2022, pp. 1--6.

\bibitem{doh2023toward}
SeungHeon Doh, Minz Won, Keunwoo Choi, and Juhan Nam,
\newblock ``Toward universal text-to-music retrieval,''
\newblock in {\em ICASSP 2023-2023 IEEE International Conference on Acoustics,
  Speech and Signal Processing (ICASSP)}. IEEE, 2023, pp. 1--5.

\bibitem{wu2023clamp}
Shangda Wu, Dingyao Yu, Xu~Tan, and Maosong Sun,
\newblock ``Clamp: Contrastive language-music pre-training for cross-modal
  symbolic music information retrieval,''
\newblock {\em arXiv preprint arXiv:2304.11029}, 2023.

\bibitem{bhargav2023music}
Samarth Bhargav, Anne Schuth, and Claudia Hauff,
\newblock ``When the music stops: Tip-of-the-tongue retrieval for music,''
\newblock in {\em Proceedings of the 46th International ACM SIGIR Conference on
  Research and Development in Information Retrieval}, 2023, pp. 2506--2510.

\bibitem{cai2020music}
Tian Cai, Michael~I Mandel, and Di~He,
\newblock ``Music autotagging as captioning,''
\newblock in {\em Proceedings of the 1st Workshop on NLP for Music and Audio
  (NLP4MusA)}, 2020, pp. 67--72.

\bibitem{manco2021muscaps}
Ilaria Manco, Emmanouil Benetos, Elio Quinton, and Gy{\"o}rgy Fazekas,
\newblock ``Muscaps: Generating captions for music audio,''
\newblock in {\em 2021 International Joint Conference on Neural Networks
  (IJCNN)}. IEEE, 2021, pp. 1--8.

\bibitem{doh2023lp}
SeungHeon Doh, Keunwoo Choi, Jongpil Lee, and Juhan Nam,
\newblock ``Lp-musiccaps: Llm-based pseudo music captioning,''
\newblock {\em arXiv preprint arXiv:2307.16372}, 2023.

\bibitem{doh2021music}
Seungheon Doh, Junwon Lee, and Juhan Nam,
\newblock ``Music playlist title generation: A machine-translation approach,''
\newblock {\em arXiv preprint arXiv:2110.07354}, 2021.

\bibitem{leszczynski2023talk}
Megan Leszczynski, Shu Zhang, Ravi Ganti, Krisztian Balog, Filip Radlinski,
  Fernando Pereira, and Arun~Tejasvi Chaganty,
\newblock ``Talk the walk: Synthetic data generation for conversational music
  recommendation,''
\newblock {\em arXiv preprint arXiv:2301.11489}, 2023.

\bibitem{dong2023musechat}
Zhikang Dong, Bin Chen, Xiulong Liu, Pawel Polak, and Peng Zhang,
\newblock ``Musechat: A conversational music recommendation system for
  videos,''
\newblock {\em arXiv preprint arXiv:2310.06282}, 2023.

\bibitem{gardner2023llark}
Josh Gardner, Simon Durand, Daniel Stoller, and Rachel~M Bittner,
\newblock ``Llark: A multimodal foundation model for music,''
\newblock {\em arXiv preprint arXiv:2310.07160}, 2023.

\bibitem{hussain2023m}
Atin~Sakkeer Hussain, Shansong Liu, Chenshuo Sun, and Ying Shan,
\newblock ``Mugen: Multi-modal music understanding and generation with the
  power of large language models,''
\newblock {\em arXiv preprint arXiv:2311.11255}, 2023.

\bibitem{tang2023salmonn}
Changli Tang, Wenyi Yu, Guangzhi Sun, Xianzhao Chen, Tian Tan, Wei Li, Lu~Lu,
  Zejun Ma, and Chao Zhang,
\newblock ``Salmonn: Towards generic hearing abilities for large language
  models,''
\newblock {\em arXiv preprint arXiv:2310.13289}, 2023.

\bibitem{liu2024music}
Shansong Liu, Atin~Sakkeer Hussain, Chenshuo Sun, and Ying Shan,
\newblock ``Music understanding llama: Advancing text-to-music generation with
  question answering and captioning,''
\newblock in {\em ICASSP 2024-2024 IEEE International Conference on Acoustics,
  Speech and Signal Processing (ICASSP)}. IEEE, 2024, pp. 286--290.

\bibitem{manco2023song}
Ilaria Manco, Benno Weck, Seungheon Doh, Minz Won, Yixiao Zhang, Dmitry
  Bodganov, Yusong Wu, Ke~Chen, Philip Tovstogan, Emmanouil Benetos, et~al.,
\newblock ``The song describer dataset: A corpus of audio captions for
  music-and-language evaluation,''
\newblock {\em arXiv preprint arXiv:2311.10057}, 2023.

\bibitem{huang2023noise2music}
Qingqing Huang, Daniel~S Park, Tao Wang, Timo~I Denk, Andy Ly, Nanxin Chen,
  Zhengdong Zhang, Zhishuai Zhang, Jiahui Yu, Christian Frank, et~al.,
\newblock ``Noise2music: Text-conditioned music generation with diffusion
  models,''
\newblock {\em arXiv preprint arXiv:2302.03917}, 2023.

\bibitem{agostinelli2023musiclm}
Andrea Agostinelli, Timo~I Denk, Zal{\'a}n Borsos, Jesse Engel, Mauro Verzetti,
  Antoine Caillon, Qingqing Huang, Aren Jansen, Adam Roberts, Marco
  Tagliasacchi, et~al.,
\newblock ``Musiclm: Generating music from text,''
\newblock {\em arXiv preprint arXiv:2301.11325}, 2023.

\bibitem{nieto2019harmonix}
Oriol Nieto, Matthew~C McCallum, Matthew~EP Davies, Andrew Robertson, Adam~M
  Stark, and Eran Egozy,
\newblock ``The harmonix set: Beats, downbeats, and functional segment
  annotations of western popular music.,''
\newblock in {\em ISMIR}, 2019, pp. 565--572.

\bibitem{reimers2019sentence}
Nils Reimers and Iryna Gurevych,
\newblock ``Sentence-bert: Sentence embeddings using siamese bert-networks,''
\newblock in {\em Proceedings of the 2019 Conference on Empirical Methods in
  Natural Language Processing and the 9th International Joint Conference on
  Natural Language Processing (EMNLP-IJCNLP)}, 2019, pp. 3982--3992.

\bibitem{team2024gemma}
Gemma Team, Thomas Mesnard, Cassidy Hardin, Robert Dadashi, Surya Bhupatiraju,
  Shreya Pathak, Laurent Sifre, Morgane Rivi{\`e}re, Mihir~Sanjay Kale,
  Juliette Love, et~al.,
\newblock ``Gemma: Open models based on gemini research and technology,''
\newblock {\em arXiv preprint arXiv:2403.08295}, 2024.

\bibitem{hu2021lora}
Edward~J Hu, Yelong Shen, Phillip Wallis, Zeyuan Allen-Zhu, Yuanzhi Li, Shean
  Wang, Lu~Wang, and Weizhu Chen,
\newblock ``Lora: Low-rank adaptation of large language models,''
\newblock {\em arXiv preprint arXiv:2106.09685}, 2021.

\bibitem{wu2023large}
Yusong Wu, Ke~Chen, Tianyu Zhang, Yuchen Hui, Taylor Berg-Kirkpatrick, and
  Shlomo Dubnov,
\newblock ``Large-scale contrastive language-audio pretraining with feature
  fusion and keyword-to-caption augmentation,''
\newblock in {\em ICASSP 2023-2023 IEEE International Conference on Acoustics,
  Speech and Signal Processing (ICASSP)}. IEEE, 2023, pp. 1--5.

\bibitem{zhuo2023video}
Le~Zhuo, Zhaokai Wang, Baisen Wang, Yue Liao, Chenxi Bao, Stanley Peng, Songhao
  Han, Aixi Zhang, Fei Fang, and Si~Liu,
\newblock ``Video background music generation: Dataset, method and
  evaluation,''
\newblock in {\em Proceedings of the IEEE/CVF International Conference on
  Computer Vision}, 2023, pp. 15637--15647.

\bibitem{song2020score}
Yang Song, Jascha Sohl-Dickstein, Diederik~P Kingma, Abhishek Kumar, Stefano
  Ermon, and Ben Poole,
\newblock ``Score-based generative modeling through stochastic differential
  equations,''
\newblock {\em arXiv preprint arXiv:2011.13456}, 2020.

\end{thebibliography}

\end{document}